\documentclass[12pt]{article}
\usepackage{graphics}
\advance\textheight by \topskip
\begin{document}
\tolerance=100000
\thispagestyle{empty}
\setcounter{page}{0}
\def\cO#1{{\cal{O}}\left(#1\right)}
\newcommand{\be}{\begin{equation}}
\newcommand{\ee}{\end{equation}}
\newcommand{\br}{\begin{eqnarray}}
\newcommand{\er}{\end{eqnarray}}
\newcommand{\ba}{\begin{array}}
\newcommand{\ea}{\end{array}}
\newcommand{\bi}{\begin{itemize}}
\newcommand{\ei}{\end{itemize}}
\newcommand{\bn}{\begin{enumerate}}
\newcommand{\en}{\end{enumerate}}
\newcommand{\bc}{\begin{center}}
\newcommand{\ec}{\end{center}}
\newcommand{\ul}{\underline}
\newcommand{\ol}{\overline}
\newcommand{\ra}{\rightarrow}
\newcommand{\sm}{${\cal {SM}}$}
\newcommand{\as}{\alpha_s}
\newcommand{\aem}{\alpha_{em}}
\newcommand{\ycut}{y_{\mathrm{cut}}}
\newcommand{\susy}{{{SUSY}}}
\newcommand{\Dir}{\kern -6.4pt\Big{/}}
\newcommand{\Dirin}{\kern -10.4pt\Big{/}\kern 4.4pt}
\newcommand{\DDir}{\kern -10.6pt\Big{/}}
\newcommand{\DGir}{\kern -6.0pt\Big{/}}
  \def\Ecm{\ifmmode{E_{\mathrm{cm}}}\else{$E_{\mathrm{cm}}$}\fi}
\def\gluino{\ifmmode{\mathaccent"7E g}\else{$\mathaccent"7E g$}\fi}
\def\photino{\ifmmode{\mathaccent"7E \gamma}\else{$\mathaccent"7E \gamma$}\fi}
\def\mgluino{\ifmmode{m_{\mathaccent"7E g}}
             \else{$m_{\mathaccent"7E g}$}\fi}
\def\taugluino{\ifmmode{\tau_{\mathaccent"7E g}}
             \else{$\tau_{\mathaccent"7E g}$}\fi}
\def\mphotino{\ifmmode{m_{\mathaccent"7E \gamma}}
             \else{$m_{\mathaccent"7E \gamma}$}\fi}
\def\ML{\ifmmode{{\mathaccent"7E M}_L}
             \else{${\mathaccent"7E M}_L$}\fi}
\def\MR{\ifmmode{{\mathaccent"7E M}_R}
             \else{${\mathaccent"7E M}_R$}\fi}
\def\lsim{\buildrel{\scriptscriptstyle <}\over{\scriptscriptstyle\sim}}
\def\gsim{\buildrel{\scriptscriptstyle >}\over{\scriptscriptstyle\sim}}
\def\MCH {$\tilde\chi_1^+$}
\def \CH{{\tilde\chi}^{\pm}}
\def \LSP{\tilde\chi_1^0}
\def \SNU{\tilde{\nu}}
\def \BARSNU{\tilde{\bar{\nu}}}
\def \MLSP{m_{{\tilde\chi_1}^0}}
\def \MCH{m_{{\tilde\chi}^{\pm}}}
\def \MCHMIN {\MCH^{min}}
\def \ET{\not\!\!{E_T}}
\def \LL{\tilde{l}_L}
\def \LR{\tilde{l}_R}
\def \MLL{m_{\tilde{l}_L}}
\def \MLR{m_{\tilde{l}_R}}
\def \MSNU{m_{\tilde{\nu}}}
\def \PI{{\pi^{\pm}}}
\def \DM{{\Delta{m}}}
\newcommand{\bQ}{\overline{Q}}
\newcommand{\ad}{\dot{\alpha }}
\newcommand{\bd}{\dot{\beta }}
\newcommand{\dd}{\dot{\delta }}
\def \CH{{\tilde\chi}^{\pm}}
\def \MCH{m_{{\tilde\chi}_1^{\pm}}}
\def \LSP{\tilde\chi_1^0}
\def \MUL{m_{\tilde{u}_L}}
\def \MUR{m_{\tilde{u}_R}}
\def \MDL{m_{\tilde{d}_L}}
\def \MDR{m_{\tilde{d}_R}}
\def \MSNU{m_{\tilde{\nu}}}
\def \MLL{m_{\tilde{l}_L}}
\def \MLR{m_{\tilde{l}_R}}
\def \mhf{m_{1/2}}
\def \MST{m_{\tilde t_1}}
\def \lum{{\cal L}}
\def \RPVC{\lambda'}
\def\tth{\tilde{t}\tilde{t}h}
\def\qqh{\tilde{q}_i \tilde{q}_i h}
\def\t1{\tilde t_1}
\def \pt{p{\!\!\!/}_T} 
\def \etm{E{\!\!\!/}_T} 
\def\lapp{\mathrel{\rlap{\raise.5ex\hbox{$<$}}
                    {\lower.5ex\hbox{$\sim$}}}}
\def\gapp{\mathrel{\rlap{\raise.5ex\hbox{$>$}}
                    {\lower.5ex\hbox{$\sim$}}}}
\newcommand{\decay}[2]{
\begin{picture}(25,20)(-3,3)
\put(0,-20){\line(0,1){15}}
\put(0,-20){\vector(1,0){15}}
\put(0,0){\makebox(0,0)[lb]{\ensuremath{#1}}}
\put(25,-20){\makebox(0,0)[lc]{\ensuremath{#2}}}
\end{picture}}
\vspace*{\fill}
\vspace{-1.2in}
\begin{center}
\end{center}
\begin{flushright}
{CMS-NOTE-2007/004\\
Accepted in European Journal of Physics C}
\end{flushright}
\begin{center}
{\Large \bf Study of Direct Photon plus Jet production in CMS Experiment at $\sqrt{s}$=14 TeV }\\[0.3cm]
\end{center}
\begin{center}
{\large Pooja Gupta\footnote{Email Address: pooja@fnal.gov}, 
B.C.Choudhary\footnote{Email Address: brajesh@fnal.gov}, S.Chatterji\footnote{Email Address: Sudeep.Chatterji@cern.ch},and  S.Bhattacharya\footnote{Email Address: bhattacharya.satyaki@gmail.com}\\[0.20cm]

Department of Physics and Astrophysics, University of Delhi,\\
Delhi, 110 007, India\\[0.20cm]}
\end{center}

\vspace{.4cm}

\begin{abstract}
{\noindent\normalsize 
We present simulation results of $\gamma$ + Jet analysis using CMS (Compact Muon Solenoid) Object-Oriented software at the Large Hadron Collider (LHC) center of mass energy $\sqrt{s}$=14 TeV. The study of direct photon production helps in validating the perturbative Quantum Chromodynamics (pQCD) and providing information on the gluon distribution in the nucleons. Direct photon processes also constitute a major background to several other Standard Model (SM) processes and signals of new physics. Thus these processes need to be understood precisely in the new energy regime. In this work, we have done a detailed study of the GEANT4 simulated $\gamma$ + jet events generated with Pythia, and the related background processes. Isolation cuts have been optimized for direct photon which improves the signal over background ratio by $\sim25\%$ as compared to previous studies done in CMS. The inclusion of a large $\Delta\phi$ cut between the photon and the leading jet at $40^0$ in the analysis leads to a further increase of $\sim15\%$ in S/B, thus giving an overall gain of $\sim42\%$ in S/B ratio.}

\end{abstract}
\vskip1.0cm
\noindent
\vspace*{\fill}
\newpage
\section*{Introduction}

Direct or prompt photons are photons that emerge directly from the hard scattering of the partons and not from secondary decays or as the radiation product of initial or final state partons. Since these photons come directly from parton-parton hard scattering, they provide a clean signature of the underlying hard scattering dynamics. Prompt photon production in hadronic interactions provides a precision test of pQCD predictions as well as information on the gluon density inside the colliding hadrons. Theoretical investigations have shown\cite{Aurenche} that the cross-section for direct photon production has a pseudorapidity ($\eta$) dependence which is sensitive to the parameterization of the gluon distributions\cite{Heath}. The LHC energy will provide an opportunity to determine gluon density in a proton in new kinematic region of $ x (2 \times 10^{-4} < x < 1.0 )$ and  $Q^{2}$ ($1.6 \times 10^{3} - 2 \times 10^{5} (GeV/c)^{2}$)\cite{Bandourin2}. The analysis of the LHC data in combination with results from the Tevatron and HERA would allow to extend the QCD analysis in $Q^{2}$ region $10^{2} < Q^{2} < 10^{5} (GeV/c)^{2}$\cite{Bandourin3}.

         The study of physics beyond the SM at the LHC also requires a complete understanding of the QCD processes. Some of the QCD processes constitute major background to other SM processes, and also to signals of new physics. Thus these processes need to be well understood precisely in the new energy regime. In this work, we have concentrated on the QCD process pp $\longrightarrow \gamma^{dir}$ + 1 jet + X, where X can be anything. This process is a major background to other Standard Model processes such as H $\longrightarrow \gamma\gamma$\cite{Higgs}, $W^{\pm} \longrightarrow \pi^\pm\gamma$\cite{wdecay}, and signatures of physics beyond the SM such as large extra dimensions\cite{LED1,LED2} and SUSY\cite{susy}.

        Inclusive direct photon production has been studied extensively by various experiments from fixed target energies\cite{D00} to Tevatron collider\cite{D0,D01,D02,D03,D04}. A recent study by the D0 experiment at Tevatron has reported on the measurment of the triple differential cross-section for $p \bar{p} \longrightarrow \gamma^{dir}$ + jet at $\sqrt{s}$=1.96 TeV\cite{AshishD0}. In this work, the photon + jet cross-sections has been compared to the next-to leading order (NLO) based on the program JETPHOX\cite{Binoth,Catani} with CTEQ6.1M PDF set of parton distribution functions\cite{cteq61pdf,cteq61pdf1}. However, both the completeness of theoretical calculations and consistency of the available data have been subject of intense discussion. The results from various experiments have shown a definite pattern of deviation from theory at low $P_{T}$ where the measured cross sections have a steeper slope. The origin of the disagreement has been attributed to the effect of initial-state soft gluon radiation which is not properly accounted for in the theoretical calculations. It has been shown that the inclusion of some additional transverse momentum ``$k_{T}$'' smearing of the partonic system due to soft-gluon emission yields better description of the data\cite{Ktsmearing,Ktsmearing1}. These effects are however expected to be negligible in the kinematic range explored in the present analysis.
       
         At the leading order, the direct photon production is defined by two QCD processes; Quark-Gluon Compton scattering, $q + g \longrightarrow \gamma^{dir}+q$ and Quark-Antiquark annihilation, $q + \overline{q} \longrightarrow \gamma^{dir} + g$. Due to abundance of low momentum fraction gluons in the proton at the LHC energy, Compton scattering becomes the dominant process contributing to the prompt photon production over most of the kinematical region.

         Unfortunately, the advantages of photon as a clean probe of parton distributions are offset by large QCD  backgrounds which are $\sim10^{3}$ to $10^{4}$ times larger than that of direct photon signal. The background contribution to direct photon is mainly caused by the events where high $P_{T}$ photons are produced in the decays of neutral mesons such as $\pi^0$, $\eta$, ${k}^{0}_{s}$ and $\omega^0$; and from events where the photons are radiated from the quark (i.e, bremsstrahlung photons in the next-to-leading order QCD subprocesses such as $qg \longrightarrow qg$, $qq \longrightarrow qq$ and $q\overline{q} \longrightarrow q\overline{q} $). Any analysis must separate true direct photons (those coming from the hard scattering) from those copiously produced in the decays of $\pi^0$, $\eta$, ${k}^{0}_{s}$, $\omega^0$ and bremsstrahlung photons emerging from  high $E_{T}$ jets. Isolation cuts imposed on reconstructed photon candidates effectively suppress these backgrounds\cite{Note200564}. The purpose of this work is to study the $\gamma$ + jet events and its background processes with full detector simulation and reconstruction, and optimize the signal over background (S/B) ratio at the LHC energy.

         In this work, we present the GEANT4 simulated results of Level-3 Triggered photons (also called High Level Trigger (HLT) photons) using CMS Software packages. The event generation for $\gamma^{dir}$+ jet signal and background processes has been done using CMKIN\cite{CMKIN}. The passage of particles through the detector geometry, showering, energy loss in the calorimeters and reconstruction of the events are modeled in the CMS simulation and reconstruction packages namely, OSCAR\cite{OSCAR} and ORCA\cite{ORCA} respectively. The simulated result has been compared with the theoretical calculations.

         The rest of the paper is organised as follows. In the next section, we discuss the physics of $\gamma$+ jet. Section 3 gives a brief description of the CMS detector relevant for the present analysis. In section 4, we discuss event generation. Section 5 describes photon isolation and its effect on signal and background. In section 6, we discuss the effect of $\Delta\phi$ cut between the photon and the jet on S/B. Statistical and systematic uncertainties are discussed in section 7. In the last section, we summarize this analysis with the conclusions.

\section*{The Physics of $\gamma$ + jet}

\begin{figure}[!Hhtb]
  \begin{center}
    \resizebox{0.5\linewidth}{0.4\linewidth}{\includegraphics{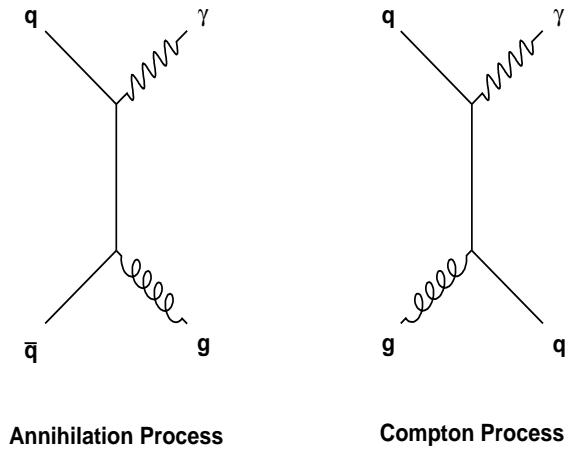}}
    \caption{Leading order processes contributing to direct photon production.}
    \label{fig:LO}
\end{center}
 \end{figure}

\begin{figure}[!Hhtb]
  \begin{center}
    \resizebox{0.5\linewidth}{0.4\linewidth}{\includegraphics{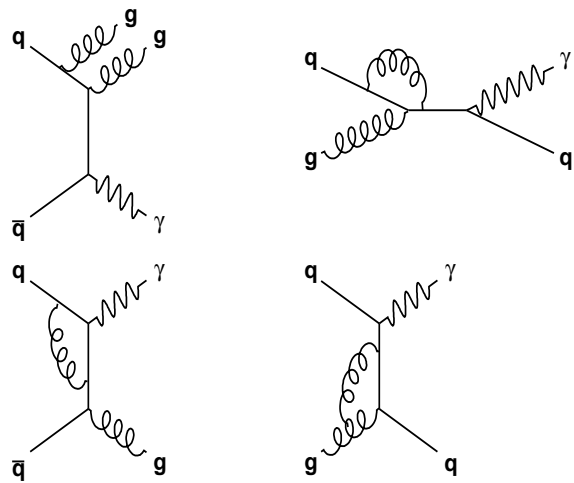}}
    \caption{NLO real diagram and corresponding virtual diagrams contributing to direct photon production.}
             \label{fig:NLO}
\end{center}
\end{figure}

 Fig.~\ref{fig:LO} shows the two leading order (LO) processes namely, ``Quark-Antiquark Annihilation'' and ``Quark-Gluon Compton Scattering'' which contributes to $\gamma$ + jet events. In quark-antiquark annihilation, a direct photon is produced along with an outgoing gluon while in quark-gluon Compton scattering, a gluon scatters from a quark producing a photon in the final state along with the scattered quark. An example of a next-to-leading order (NLO) real correction and corresponding virtual corrections (which cancel the soft and collinear divergences arising from the real radiation) are shown in fig.~\ref{fig:NLO}.  
 
If the initial parton energies are equal, the photon and the jet are produced back to back ($\eta \approx 0$). In case of large imbalance between initial parton energies, the photon and the jet tend to be produced at small angles, either both forward or backward. Since the gluon momentum distribution increases monotonically at lower momentum fraction values\cite{parton}, high $\eta$ photons are likely to come from gluon Compton scattering process rather than annihilation process.

\begin{figure}[!Hhtb]
  \begin{center}
    \resizebox{0.5\linewidth}{0.4\linewidth}{ \includegraphics{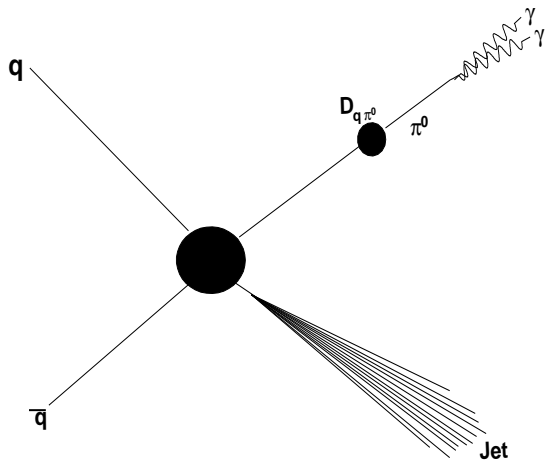}}
 \caption{Background contributions to Direct Photon Events from EM jets.}
   \label{fig:dirph-bkjets2}
\end{center}
\end{figure}

\begin{figure}[!Hhtb]
  \begin{center}
    \resizebox{0.5\linewidth}{0.4\linewidth}{\includegraphics{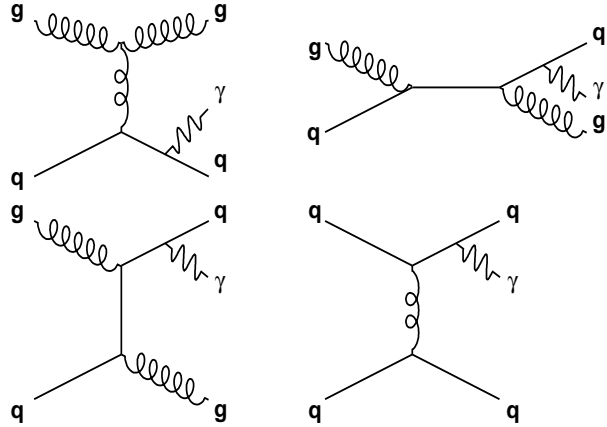}}
\caption{Background contributions to Direct Photon Events from final-state radiation processes.}
   \label{fig:dirph-bkanam1}
\end{center}
\end{figure}

         The main background to the direct photon signal comes from jet fluctuations. While most of the jets consist of many particles, thus easily distinguishable from a single photon, a small fraction ($\sim10^{-3}$ - $10^{-4}$) of jets fragments in such a way that  a single particle carries most of the momentum of the parent parton. The lightest and therefore most commonly produced neutral meson, the $\pi^{0}$ decays into two photons with a branching ratio of $\sim99\%$. Photons produced  in decays of high $P_T$ $\pi^{0}$ (and other neutral mesons) are very close to each other in the electromagnetic calorimeter forming a single shower thus mimicking a direct photon. Therefore, non-direct photons are common in hadronic collisions. Since jet production rates are $\sim10^{3}$ times larger than that of direct photons (depends on $|\eta|$  and $P_{T}$), the number of jets which fragment to a single meson and then decay to photons, referred as electromagnetic (EM) jets, are comparable to the level of direct photons. The background to direct photon + jet can also come from photons produced in initial and final state radiation. In these cases, the photon is not produced directly from the interaction vertex, and therefore, is not really a “direct” photon. In order to reduce the background contribution, an isolation criterion is imposed on the photon: typically a cone of size R (in pseudorapidity $(\eta)$ and azimuthal angle $(\phi)$) around the photon is required to have less than a certain amount of energy. Since a photon produced through bremsstrahlung tends to be collinear with the parent quark (and thus the jet resulting from the quark's subsequent fragmentation), the requirement that the photon be isolated removes all but very large angle radiations. Fig. ~\ref{fig:dirph-bkjets2} and ~\ref{fig:dirph-bkanam1} show major backgrounds to direct photons, namely, the EM jets and the bremsstrahlung.

\section*{The CMS Detector}

 The Compact Muon Solenoid (CMS) is a general purpose detector with 4$\pi$ steridian coverage for  efficient detection and precise energy/momentum measurement of electrons, photons, muons, hadrons (jets) and missing transverse energy. From inside to outside, it comprises of a pixel detector, a silicon microstrip tracker system, a lead tungstate crystal electromagnetic calorimeter (ECAL), a sampling hadronic calorimeter (HCAL), a 4 Tesla soleniodal magnet and the muon detectors. 

The innermost component of the tracking system is the highly segmented silicon pixel detector which consists of 3 layers of barrel complemented with 2 disks on each side in the endcaps. It consists of 66 million pixels of size 100 $\mu$m $\times$ 150 $\mu$m yielding a three dimensional spatial resolution of  $\sim$10 $\mu$m in the r-$\phi$ and $\sim$20 $\mu$m in the z-coordinate, which will play a crucial role in the vertex identification and track reconstruction. The following part of the tracking system consists of silicon microstrip detector modules arranged in ten concentric layers in barrel and nine disks in the endcaps providing a single point resolution between 23-52 $\mu$m in r-$\phi$ and 230-530 $\mu$m in z. It provides the necessary lever arm for accurate momentum measurement and improves the precision of vertex measurement. The tracker has geometrical coverage upto $|\eta|=2.5$. The momentum resolution $\Delta P_{T}/P_{T}$ of high $P_{T}$ charged particles in the tracker is $\sim (15(P_{T}/TeV)+0.5)\%$ for $|\eta| < 1.6$ and becomes $\sim (60(P_{T}/TeV)+0.5)\%$ as $|\eta|$ approaches 2.5\cite{TrackerTDR,TrackerAdd,PhyTDR1}.

The electromagnetic calorimeter is a hermetic, homogeneous calorimeter consisting of 61,200 lead tungstate $({PbWO}_{4})$ crystals mounted in central part corresponding to $0<|\eta|<1.479$, and 7324 crystals mounted in each of the two endcaps in the range of $1.479<|\eta|<3.0$. The crystals are 23 cm long, have short radiation length($X_{0}$ =0.89 cm) of Moliere radius(21.9 mm) and correspond to $\sim$26$X_{0}$, thereby containing more than 99$\%$ of the shower energy. The granularity of the calorimeter is $\Delta\eta$ x $\Delta\phi  = 0.0175$ x $0.0175$ in the barrel and upto $\Delta\eta$ x $\Delta\phi  = 0.05$ x $0.05$ in the endcaps. The resolution for single photons with energies of 25-500 GeV can be parameterized as $\sigma_{E}/E$ = $3.6\%/\sqrt{E} \oplus 124/E \oplus 0.26\%$. An additional Pb-Si preshower detector would be installed in front of the  endcaps covering $1.65<|\eta|<2.61$ for efficient rejection of $\gamma/\pi^{0}$\cite{PhyTDR1,ECALTDR}. 

The sampling hadronic calorimeter is made up of plastic scintillator tiles inserted between brass absorber plates and extends upto $|\eta|<3.0$. The barrel (endcaps) consists of 5 cm (8 cm) thick brass plates interleaved with 3.8 mm thick scintillator tiles. The barrel calorimeter has a depth of 79 cm corresponding to 5.15 hadronic interaction lengths $(\lambda_{I})$ and a segmentation of $\Delta\eta$ x $\Delta\phi  = 0.087$ x $0.087$. Outside of the soleniod, an additional layer of 10 mm thick scintillator called outer hadron calorimeter covering the $|\eta|<1.26$ region, is placed to ensure sampling of hadronic shower with $\sim$ 11 $\lambda_{I}$. Hermecity of the  HCAL is extended to $|\eta|<5.0$ by adding forward detector made up of steel absorbers and embedded radiation hard quartz fibers, which provide a fast collection of Cerenkov light. The resolution of the HCAL ranges from $\sigma_{E}/E$ = $65\%/ \sqrt{E}$  $\oplus  5\%$ in the barrel (at $\eta=0$) and $\sigma_{E}/E$ = $83\%/\sqrt{E} \oplus 5\%$ in the endcaps, to $\sigma_{E}/E$ = $100\%/\sqrt{E} \oplus 5\%$ in the forward region\cite{PhyTDR1,HCALTDR}.

The muon detector consists of drift tubes in the barrel $(|\eta|<1.2)$ and cathode strip chambers in the endcaps $(1.2<|\eta|<2.4)$ for precision track measurement. A set of resistive plate chambers are used in both parts of the system for resolving individual bunch crossing time. The spatial resolution of the muon system varies between 50 to 200 $\mu$m and standalone momentum resolution is at most 15$\%$ for a 10 GeV $P_{T}$ muon and 40$\%$ at 1 TeV\cite{PhyTDR1,MuonTDR}.

\section*{Event Generation}

         At the LHC center of mass energy, the cross-sections for the channels pp $\longrightarrow \gamma $+ jet and pp $\longrightarrow $ jets are very large, thus making it nearly impossible to simulate direct photon signal and its background processes equivalent to an integrated luminosity of $\sim1 fb^{-1}$. In order to achieve this, a selection based on kinematic information of the particle is performed at the generator level which allows only those events to be fully simulated that are likely to pass the analysis criteria. In our study we have used $|{\eta}_{EMCAL}|<2.6$ to match the preshower coverage. The CMS experiment has optimized the L1 trigger $P_{T}$ cut for a L1 isolated single photon to be 23 GeV, while that for the HLT photon it is 80 GeV at a luminosity of $2 \times {10}^{33}{cm}^{-2}{s}^{-1}$\cite{PhyTDR1}. For the signal, a pre-selection has been used at the generator level which retains only those events having a direct photon with  $P^{\gamma}_{T} > 70$ GeV and $|\eta^{\gamma}| < 2.8$, which are well below the final analysis cut of $P^{\gamma}_{T} > 80$ GeV and $|\eta^{\gamma}| < 2.6$. This preselection helps in discarding those events before simulation which are unlikely to pass geometrical acceptance cut. The L1 trigger efficiency for the signal has been estimated to be above $99\%$ over the entire kinematic range\cite{PhyTDR1}. 

For the backgrounds, we have used Pythia generated QCD dijet events. The samples $(eg03\_jets\_1e)$ have been preselected at the generator level with an enhanced L1 trigger efficiency\cite{Note200564}. In this pre-selection, the first step is to look for the seed particles of electromagnetic objects like photons, electrons and positrons, which have $P_{T} > 5$ GeV and $|\eta| < 2.7$. Candidate electromagnetic calorimeter trigger tower energies are then estimated by adding energies of all electromagnetic particles found within a cone of $\Delta\eta<$0.09 and $\Delta\phi<$0.09 from the seed. Trigger tower candidates that lie within $\Delta\eta<0.2$ and $\Delta\phi<0.2$ from each other are suppressed and only those with the highest $P_{T}$ are retained. The L1 single photon electromagnetic trigger is simulated by requiring that one such candidate has transverse energy larger than 20 GeV. 
  
   Signal and background events have been reconstructed using ORCA version 8.7.3. In the electromagnetic calorimeter, the Hybrid and Island clustering algorithms\cite{Note2001034} were used to construct basic clusters in the barrel and the endcap regions respectively. A supercluster is a cluster of the basic clusters. A photon candidate is associated with each reconstructed supercluster. Events were generated in two different $P_{T}$ bins: $50$ GeV $<P_{T}<170$ GeV, and $P_{T}>170$ GeV.  

\begin{figure}[!Hhtb]
  \begin{center}
    \resizebox{0.65\linewidth}{0.5\linewidth}{\includegraphics{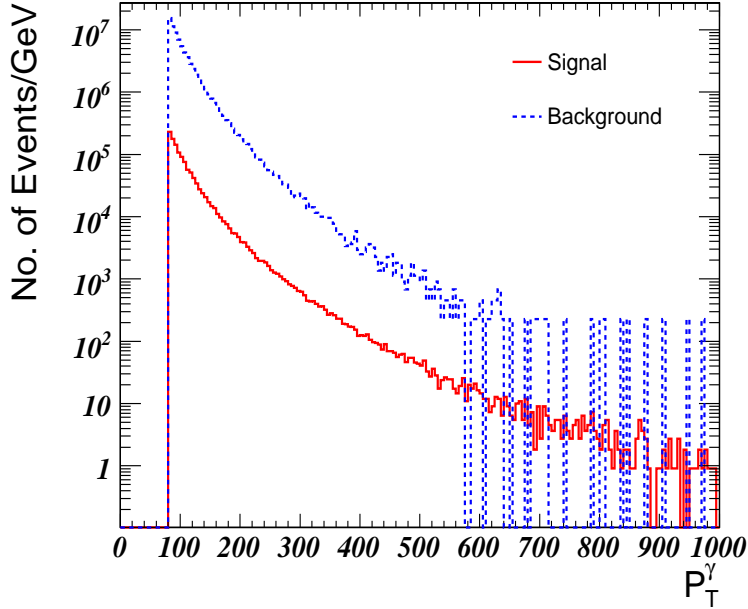}}
\caption{Number of Events/GeV for the $\gamma$ + jet signal and its backgrounds for an integrated luminosity of 1 $fb^{-1}$. No isolation requirements have been applied on the photon.}
\label{fig:numberofeventswithoutisolation}
\end{center}
\end{figure}

    The cross-sections, number of events and corresponding integrated luminosity for signal and background samples are shown in Table ~\ref{table:cross-section}. The samples have been generated using Pythia\cite{pythia}, simulated  and reconstructed using OSCAR and ORCA. We require that the leading photon should have $P_{T}^{\gamma}>$ 80 GeV and $|\eta^{\gamma}| < 2.6$. Fig.~\ref{fig:numberofeventswithoutisolation} shows the number of events/GeV after $P_{T}$ and $\eta$ requirements calculated as a function of ${P}^{\gamma}_{T}$ for the $\gamma$ + jet signal and its background for an integrated luminosity of 1 $fb^{-1}$.

{\small
\begin{table*}[!Hhtb]
\begin{center}
\begin{tabular}{|c|c|c|c|c|c|}
\hline

\textbf{Data} & \textbf{$P_{T}$}  & \textbf{$\sigma$ generated}  &\textbf{$\sigma$ preselected} & \textbf{Number of} & \textbf{Integrated}\\
\textbf{Samples}& \textbf{(in GeV)}  & \textbf{(in $pb$)}    &\textbf{(in $pb$)}  & \textbf{Simulated}   & \textbf{Luminosity}\\
 &&& &\textbf{Events} &\textbf{(in $pb^{-1}$)}\\
\hline

$\gamma$ + jet &50-120  & 8.53E+03    & 1.82E+03  & 30000  & 16.5\\ & & && &\\
$\gamma$ + jet &120-170  & 2.77E+02    & 2.35E+02 & 12740  & 54.2\\  &&&& &  \\
$\gamma$ + jet &170-230  & 6.76E+01    & 6.19E+01 & 11918  & 192.5\\ &&& & &\\
$\gamma$ + jet &230-300  & 1.90E+01    & 1.83E+01 & 11531   & 630.9\\ &&& & &\\
$\gamma$ + jet & $>$ 300  & 9.07E+00    & 8.98E+00   & 9887  & 1212.3\\ &&& & &\\
\hline
eg03\_jets\_1e\_50170  &50-170  & 2.43E+07    & 4.35E+06 & 2314853  & 0.53\\&&&  & &\\
eg03\_jets\_1e\_170up  & $>$ 170  & 1.34E+05    & 1.07E+05 & 473668  & 4.4\\  &&&& &\\
\hline
\end{tabular}
\end{center}
\caption{ Cross-sections, number of simulated events, and equivalent integrated luminosity for $\gamma$ + jet events and its background samples.
\label{table:cross-section}}
\end{table*}
}

\normalsize

\section*{Photon Isolation}

As shown in fig.~\ref{fig:numberofeventswithoutisolation}, jets misidentified as photons give a very large background to the direct photon signal. Isolation cuts have been optimized and applied to reject photons coming from jets by detecting other fragments of the jet which accompany the photon. 

\subsection*{Tracker Isolation}
Direct photons produced in hard scattering are not accompanied by any charged particles, whereas in cases where a jet fragments into a $\pi^{0}$ mimicing a photon or the jet bremms a  photon, the electromagnetic object is usually accompanied by tracks of low momentum charged hadrons. Thus the tracker isolation is an important parameter in differentiating the signal from the background events. The tracker isolation criteria is based on the number of charged particle tracks above a certain $P_{T}$ threshold calculated in a cone R (R=$\sqrt{\Delta\eta^{2}+\Delta\phi^{2}}$) around the photon candidate. The algorithm contains following three parameters:
\linespread{0.7}
\begin{enumerate}
\item Cone size R: The size of the cone R around the photon candidate, in which the number of charged tracks is counted.
\linespread{0.7}
\item $P_{T}$ threshold $(P^{Thres}_{T})$: Only tracks greater than threshold $P_{T}$, $P^{Thres}_{T}$, are  considered for photon isolation.
\linespread{0.7}
\item Number of tracks threshold ($N^{Thres}_{TK}$): If the number of tracks in cone R with track $P_{T} > P^{Thres}_{T}$ is larger than $N^{Thres}_{TK}$, then the photon candidate is considered non-isolated, otherwise it is considered as an isolated photon.
\end{enumerate}

\begin{figure*}[!Hhtb]
  \begin{center}
    \resizebox{0.45\linewidth}{0.35\linewidth}{\includegraphics{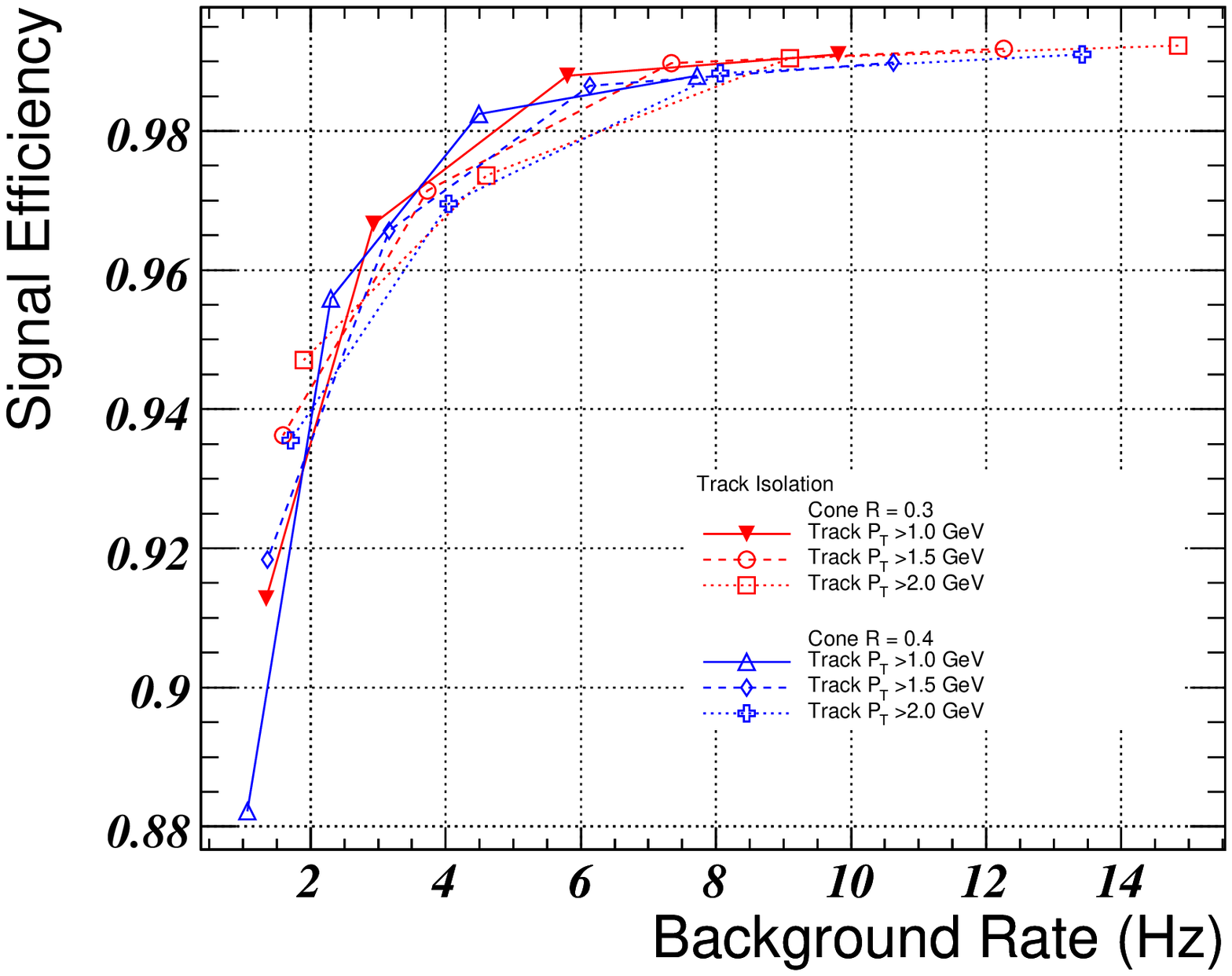}}\thinspace
\resizebox{0.45\linewidth}{0.35\linewidth}{\includegraphics{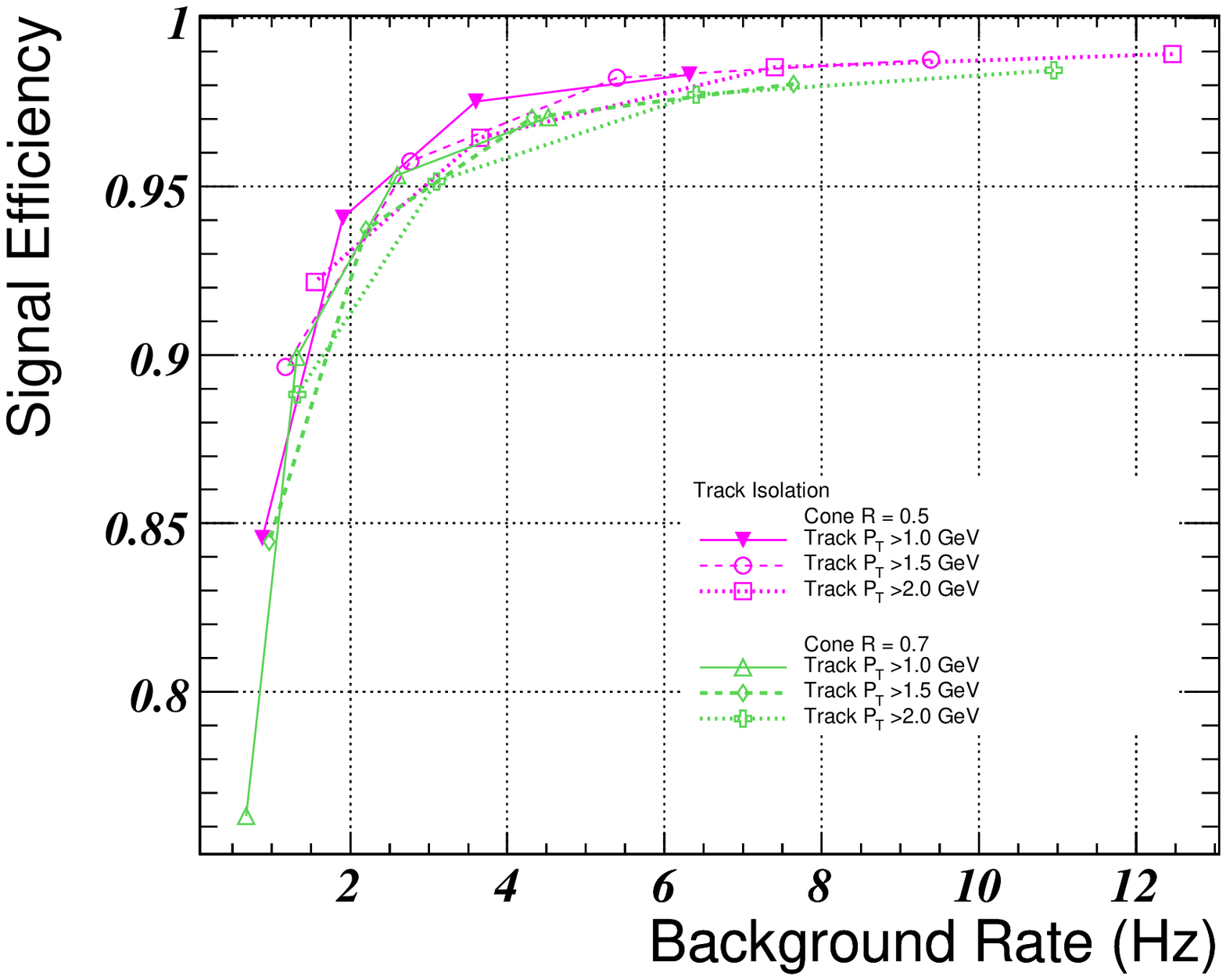}} \\
\textbf{ (a)  \qquad  \qquad  \qquad  \qquad  \qquad \qquad  \qquad  \qquad       (b)}
\caption{Performance of the tracker isolation variables for different track $P^{Thres}_{T}$ in the barrel for cone sizes R =0.3 and 0.4 (a); R = 0.5 and 0.7 (b).}
\label{fig:Trksig_eff_bkg_rate_barrel}
\end{center}
\end{figure*}

\begin{figure*}[!Hhtb]
  \begin{center}
    \resizebox{0.45\linewidth}{0.35\linewidth}{\includegraphics{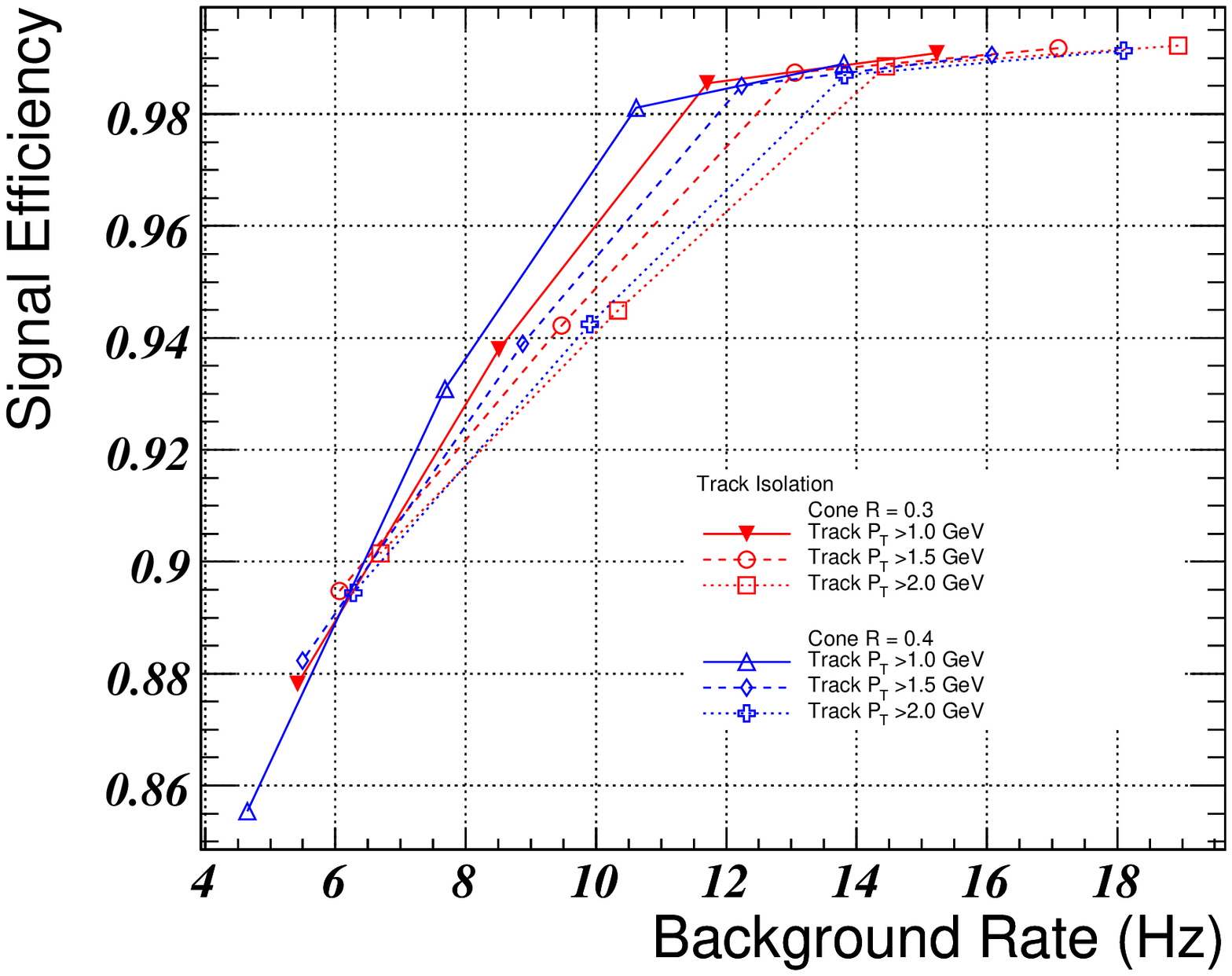}}\thinspace 
\resizebox{0.45\linewidth}{0.35\linewidth}{\includegraphics{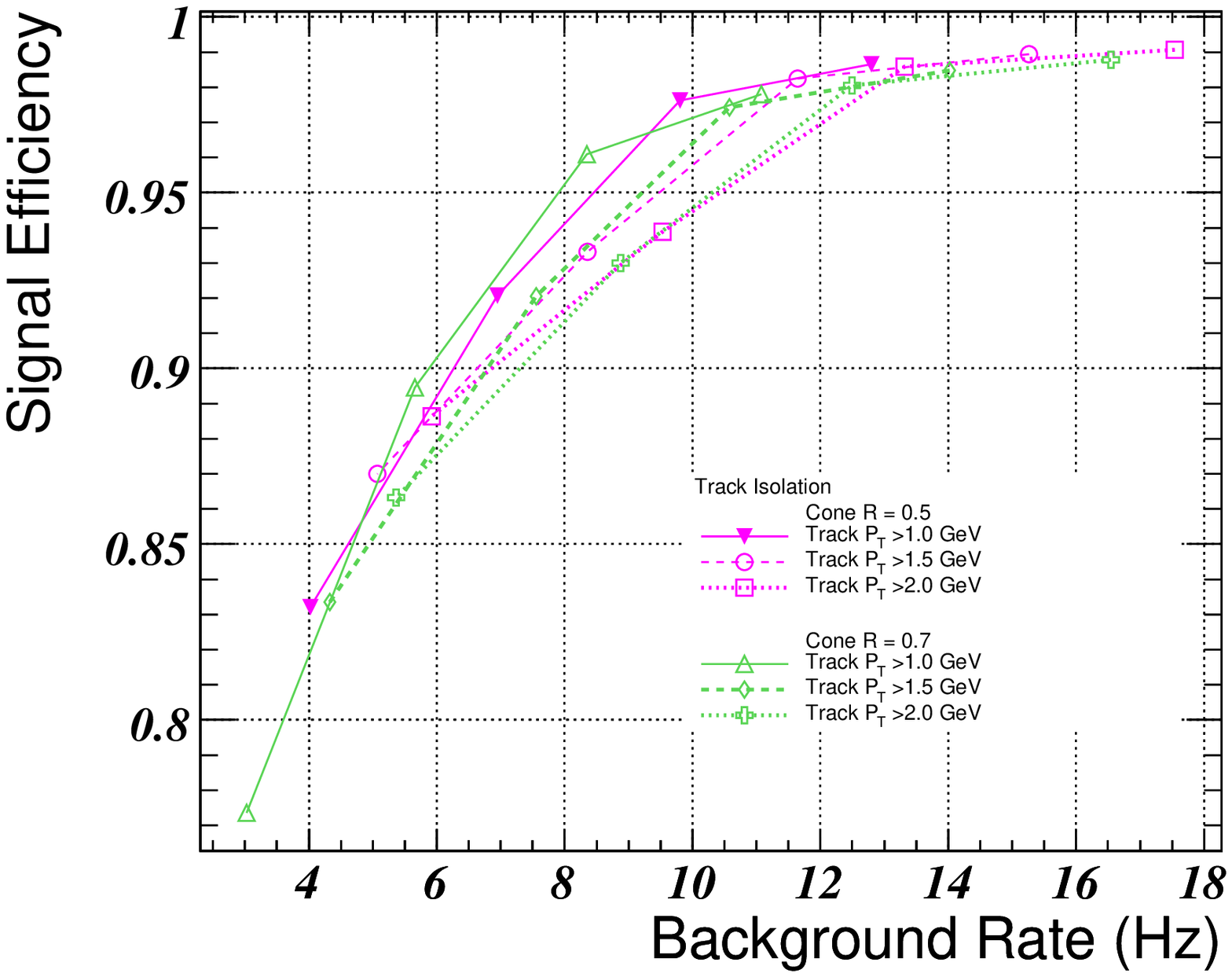}}  \\
\textbf{ (a)  \qquad  \qquad  \qquad  \qquad  \qquad \qquad  \qquad  \qquad       (b)}
\caption{Performance of the tracker isolation variables for different track $P^{Thres}_{T}$ in the endcaps for cone sizes R =0.3 and 0.4 (a); R = 0.5 and 0.7 (b). }
\label{fig:Trksig_eff_bkg_rate_endcap}
\end{center}
\end{figure*}

We have done a detailed study of the effect of $P_{T}$ threshold of the track $(P^{Thres}_{T})$ and the number of allowed tracks $(N^{Thres}_{TK})$ inside the isolation cone (R), on signal efficiency and background rates. The study has been done separately for the barrel and the endcaps. The results for the barrel and the endcaps are shown in fig.~\ref{fig:Trksig_eff_bkg_rate_barrel} and fig.~\ref{fig:Trksig_eff_bkg_rate_endcap} respectively. Now limiting the discussion to fig.~\ref{fig:Trksig_eff_bkg_rate_barrel}(a) only, the six curves show the signal efficiency as a function of background rate for three different values of track $P_{T}$ threshold (1.0 GeV, 1.5 GeV and 2.0 GeV) for each of the two cone sizes (R=0.3 and R=0.4). From left to right, the five points on each of these curves show the change in signal efficiency and the background rate as the number of tracks allowed in the vicinity of the photon increases from 0 to 4. The background rate is found to be very sensitive to the number of tracks allowed inside the cone. For example, if the number of allowed track changes from 0 to 1, the signal efficiency increases by 4-10$\%$ while the background rate goes up by a factor of $\sim2$ in the barrel and $\sim1.5$ in the endcaps. Hence, for the maximum background rejection, $N^{Thres}_{TK}$ is fixed to zero. For a fixed cone size, both signal efficiency and background rate increase with increase in track $P_{T}$ threshold. Due to larger tracker inefficiencies the performance of the tracker variables is not as efficient in endcaps. Different values of $P^{Thres}_{T}$ and cone size have been taken into account to estimate the optimized S/B ratio.

\begin{figure*}[!Hhtb]
  \begin{center}
    \resizebox{0.45\linewidth}{0.35\linewidth}{\includegraphics{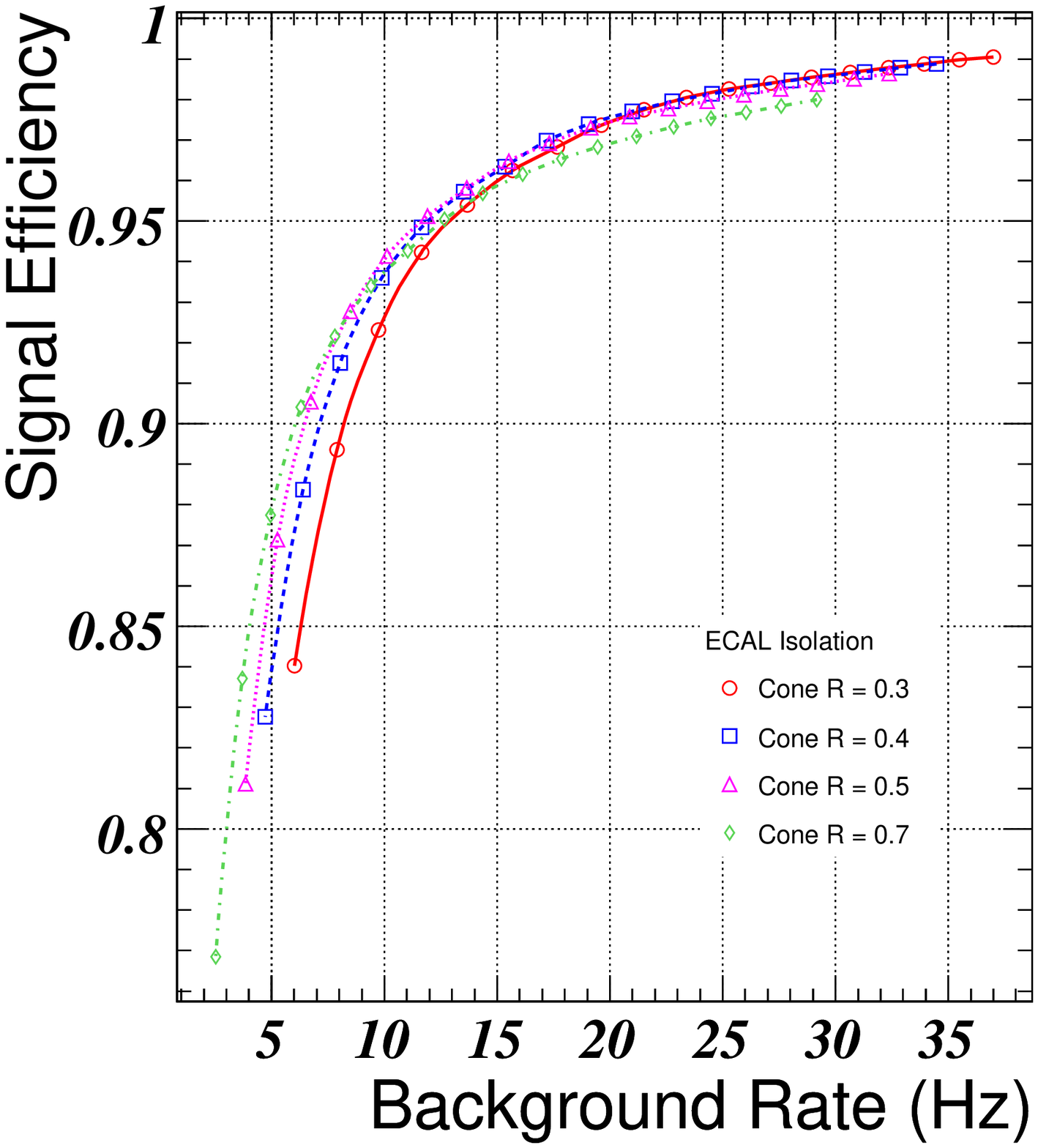}}\thinspace 
\resizebox{0.45\linewidth}{0.35\linewidth}{\includegraphics{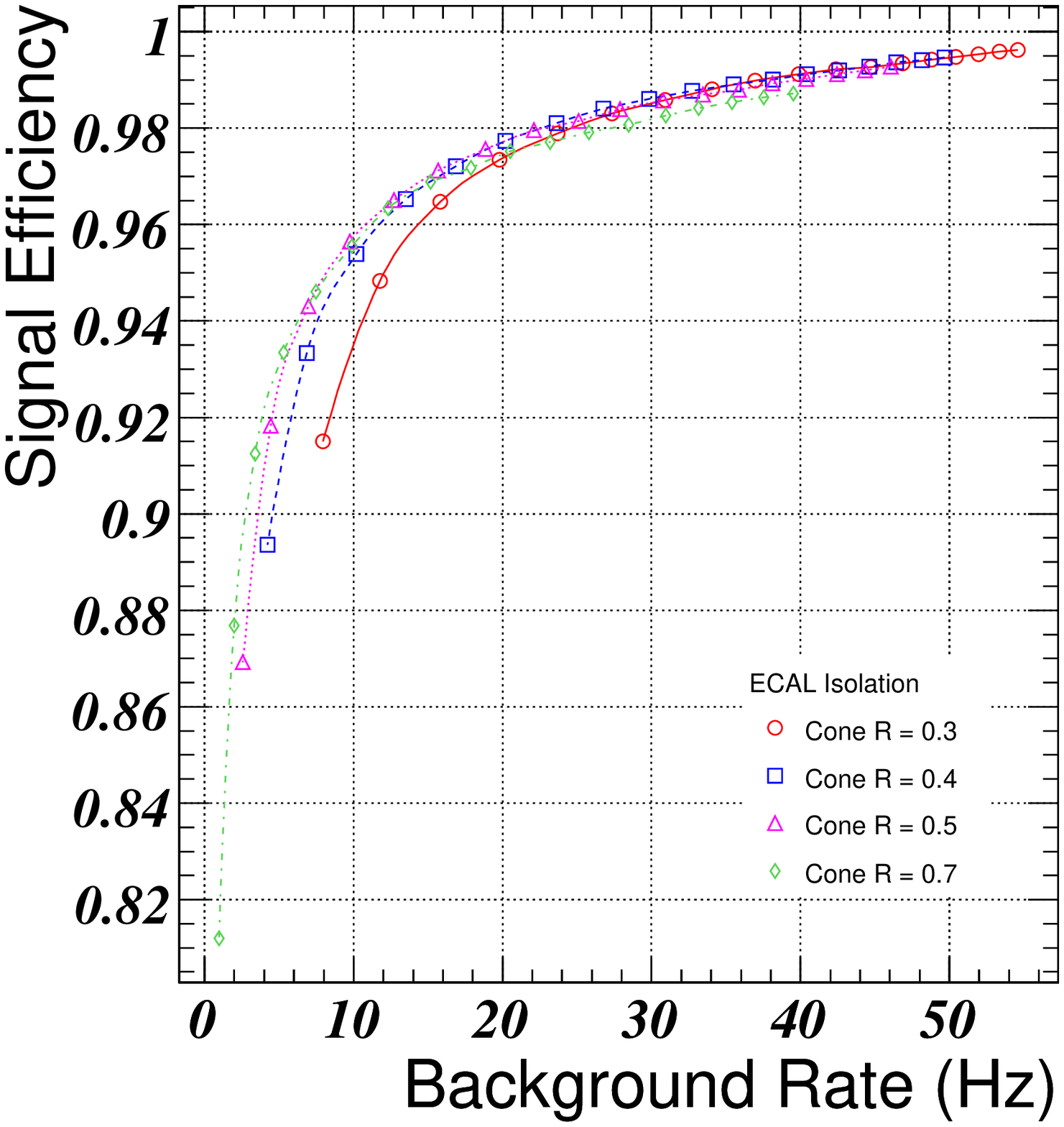}} \\
\textbf{ (a)  \qquad  \qquad  \qquad  \qquad  \qquad \qquad  \qquad \qquad     (b)}
\caption{Performance of the variable $E_{T ECAL}$ for various cone sizes R =0.3, 0.4, 0.5 and  0.7 for the barrel (a); and for endcaps (b). }
\label{fig:ECALsig_eff_bkg_rate}
\end{center}
\end{figure*}

\begin{figure*}[!Hhtb]
  \begin{center}
    \resizebox{0.45\linewidth}{0.35\linewidth}{\includegraphics{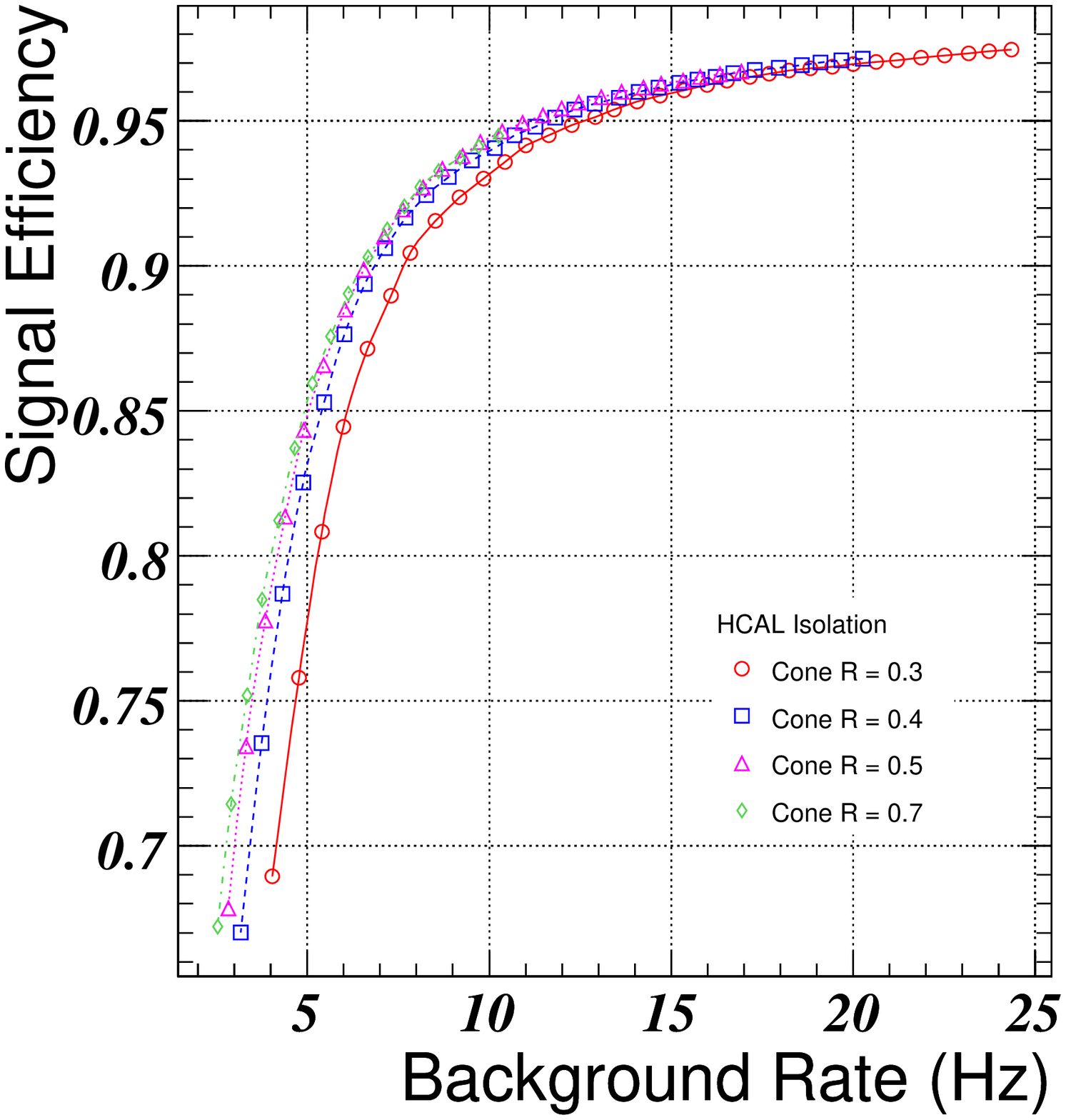}}\thinspace 
 \resizebox{0.45\linewidth}{0.35\linewidth}{\includegraphics{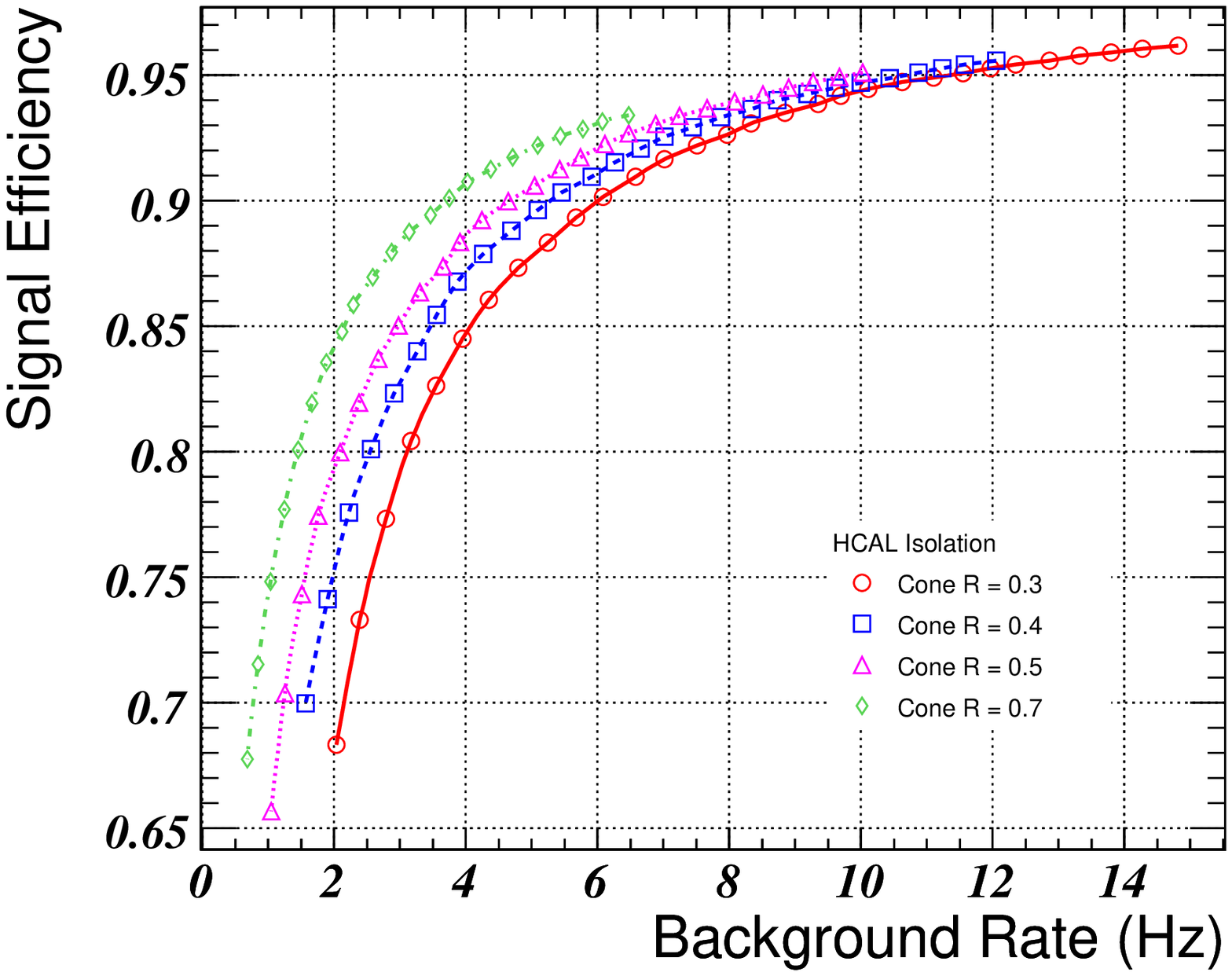}}\\
\textbf{ (a)    \qquad \qquad  \qquad \qquad  \qquad  \qquad \qquad  \qquad  (b)}
\caption{Performance of the variable $E_{T HCAL}$ for various cone sizes R =0.3, 0.4, 0.5 and  0.7 for the barrel (a); and for endcaps (b). }
\label{fig:HCALsig_eff_bkg_rate}
\end{center}
\end{figure*}

\subsection*{Electromagnetic Calorimeter Isolation}

       The ECAL isolation is based on the sum of transverse energies deposited in the basic clusters in cone of size R in $\eta-\phi$ plane around a photon candidate. The basic clusters that belong to the photon candidate supercluster are not counted as a part of the sum. If the sum of transverse energies ($\Sigma E_{T ECAL}$) is below a certain threshold value ($E^{Thres}_{T ECAL}$), the photon candidate is considered isolated, otherwise it is considered non-isolated.

    The effect of making the values of $E^{Thres}_{T ECAL}$ increasingly stringent on the signal efficiency as well as on background rate for different cone sizes R for the barrel and the endcaps can be seen in fig. ~\ref{fig:ECALsig_eff_bkg_rate} (a) and (b) respectively. From right to left, the value of $E^{Thres}_{T ECAL}$ is gradually decreased from a very large value in steps of 0.5 GeV to yield the curves. Just to illustrate the nature of dependence of the signal efficiency and background rates, the plots shown have variation in $E^{Thres}_{T ECAL}$ from 17 GeV to 1 GeV.  The stringent selection cut on $\Sigma E_{T ECAL}$ provides a good background rejection. A tighter $E_{T ECAL}$ cut in the endcaps as compared to the barrel region can further enhance background rejection.

\subsection*{Hadron Calorimeter Isolation}

Usually the hadronic activity around the background photons tends to be much more than around the direct photons. Thus, the HCAL isolation can also be used for photon selection. For isolation in HCAL, the sum of transverse energies of all the particles depositing energy in a cone of size R around a photon candidate is calculated. If the sum of transverse energies of the hadronic particles, $\Sigma E_{T HCAL}$ is below a certain threshold value ($E^{Thres}_{T HCAL}$), the photon candidate is considered isolated. Since the information from the charged mesons and hadrons is used for tracker isolation, HCAL  isolation offers partially redundant information. 

 Fig. ~\ref{fig:HCALsig_eff_bkg_rate} (a) and (b) show the effect of variation in the $E^{Thres}_{T HCAL}$ on the signal efficiency and the background rate for different cone sizes for the barrel and the endcaps respectively. From right to left the $E^{Thres}_{T HCAL}$ is gradually decreased from a very large value in steps of 0.5 GeV to yield the curves. The HCAL isolation variable offers a better background rejection in the endcaps.  

\subsection*{Combination of the Detectors}

\begin{figure}[!Hhtb]
  \begin{center}
    \resizebox{0.65\linewidth}{0.5\linewidth}{\includegraphics{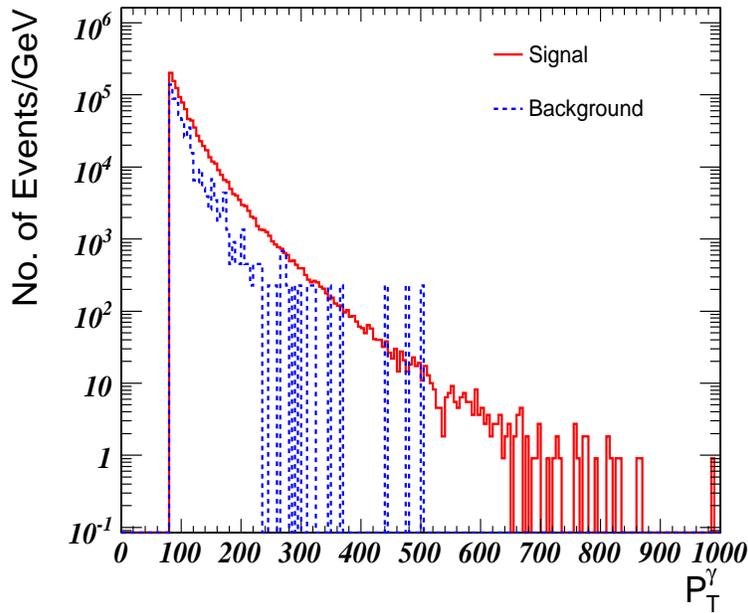}}
 \caption{Number of Events/GeV for the $\gamma$ + jet signal and its background for an  integrated luminosity 1 $fb^{-1}$ after applying the selection cut C.}
\label{fig:numberofevents}
\end{center}
\end{figure}

After studying the isolation parameters separately, the task is to combine their effects such that the S/B ratio is maximized while retaining a very high signal efficiency. This study has been done for the various detector combinations (for example: Tracker + HCAL, Tracker + ECAL, and Tracker + ECAL + HCAL) for different threshold parameters. We observe that a better reduction in the background rate without compromising the signal efficiency can be achieved by combining the isolation requirements from various detectors. Some combinations which are found to yield better signal to background ratio are shown in Table~\ref{table:Cmp_Selection}. Table~\ref{table:Cmp_Rate} shows the signal efficiency, signal and background rates, and S/B values for respective selection cuts mentioned in Table~\ref{table:Cmp_Selection}. It should be noted that the selection cut A has been used in the results mentioned in CMS Physics TDR-I\cite{PhyTDR1}. With selection cut C, the background rate is further reduced by $\sim22\%$ with only $1\%$ loss in the signal efficiency leading to a gain of $\sim26\%$ in the S/B ratio. Fig.~\ref{fig:numberofevents} shows the number of events/GeV for an integrated luminosity of 1 $fb^{-1}$ calculated as a function of photon  $P_{T}$ for the $\gamma$ + jet signal and its background for selection criteria C. The effect of the isolation requirement on the number of events/GeV can be observed by comparing fig.~\ref{fig:numberofeventswithoutisolation} and fig.~\ref{fig:numberofevents}.

\begin{table*}[htb]
\begin{center}
\begin{tabular}{|c|c|c|c|c|c|}
\hline
                                                                             
Selection & Cone Size R  & Track$P^{Thres}_{T}$ & $E^{Thres}_{T ECAL}$  &$E^{Thres}_{T HCAL}$  & $E^{Thres}_{T HCAL}$ \\
 Cut        &        &(GeV)     &(GeV)        &(in Barrel)(GeV)      &(in Endcaps)(GeV)                        \\
\hline
A	&0.3 &1.5   &1.5  &6.0  &4.0             \\
B        &0.3 &1.0  &1.5   &6.0 &5.0             \\
C        &0.4  &1.5 &2.0  &7.0  &5.0                     \\
D        &0.4 &1.5    &2.0   &6.5  &5.0                 \\
E        &0.5  &1.5    &2.5  &8.0   &6.0               \\

\hline 
\end{tabular}
\end{center}
\caption{ Selection Cuts corresponding to the different values of isolation parameters threshold.
\label{table:Cmp_Selection}}
\end{table*}

\begin{table*}[htb]
\begin{center}
\begin{tabular}{|c|c|c|c|c|}
\hline
                                                                             
\textbf{Selection} & \textbf{Signal}  & \textbf{Signal}  &\textbf{Background}  & \textbf{S/B}\\
\textbf{Cut} & \textbf{Efficiency}  & \textbf{Rate (Hz)}  &\textbf{Rate (Hz)}  & \textbf{Ratio}\\

\hline
A	&0.76		& 2.12   & 1.40  & 1.52\\
B        & 0.76		&2.10    & 1.26  &1.66\\
\textbf{C}        &\textbf{0.74}          &\textbf{2.09}    & \textbf{1.09}   &\textbf{1.92}\\
D        &0.73          &2.07    &1.06    &1.94\\
E        &0.70          &1.99    &0.84    &2.37\\

\hline 
\end{tabular}
\end{center}
\caption{ Signal efficiency, signal and background rates for various selection cuts mentioned in Table~\ref{table:Cmp_Selection} for low luminosity (L = $2\times{10}^{33}{cm}^{-2}{s}^{-1}$) at $\sqrt{s}$=14 TeV.
\label{table:Cmp_Rate}}
\end{table*}

\begin{figure}[!Hhtb]
  \begin{center}
    \resizebox{0.65\linewidth}{0.5\linewidth}{\includegraphics{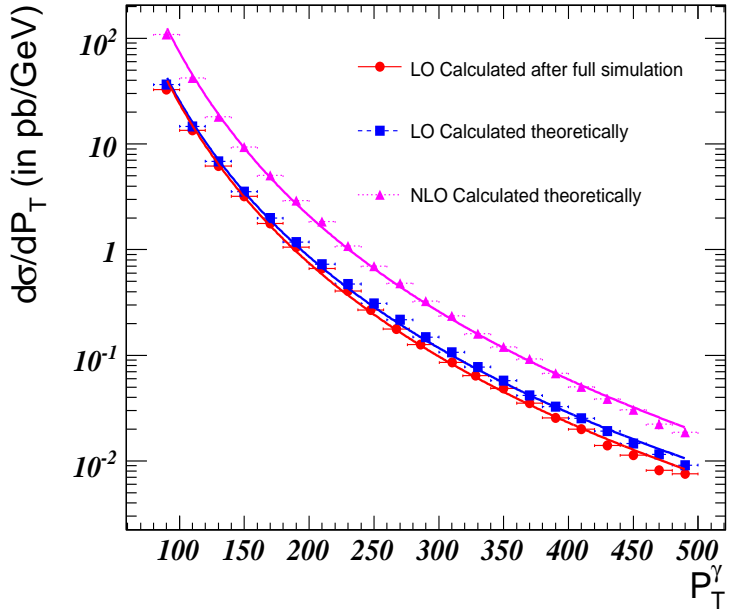}}
 \caption{The LO (after full simulation) and theoretical LO and NLO calculations for the cross-section of $\gamma$ + jet events as a function of $P_{T}^{\gamma}$. The calculated points have been complemented by lines fitting the corresponding cross sections.}
\label{fig:Diff_crossx}
\end{center}
\end{figure}

Fig.~\ref{fig:Diff_crossx} shows the the predictions for the differential cross-section of the $\gamma$ + jet events as a function of $P^{\gamma}_{T}$ compared with theoretical LO and NLO calculations\cite{Owens1}. The error bars on the x-axis represents the bin width. The differential cross-section calculated from the simulated data is represented as LO (calculated after full simulation) in the above mentioned figure. To evaluate it, the cross-section from Pythia was generated with CTEQ5L parton distribution function and renormalization scale $\mu = P_{T}$. Various factors like geometrical acceptance, efficiency of the analysis cuts and luminosity have been taken into account. Theoretical LO and NLO calculations are provided by J. Owens\cite{Owens}. The results are found to be in good agreement with the theoretical LO calculation. As expected the NLO QCD contribution is higher than the LO in the whole $P_{T}$ range under analysis. 

\section*{ $\Delta\phi$ Cut }
\begin{figure}[!Hhtb]
  \begin{center}
    \resizebox{0.65\linewidth}{0.5\linewidth}{\includegraphics{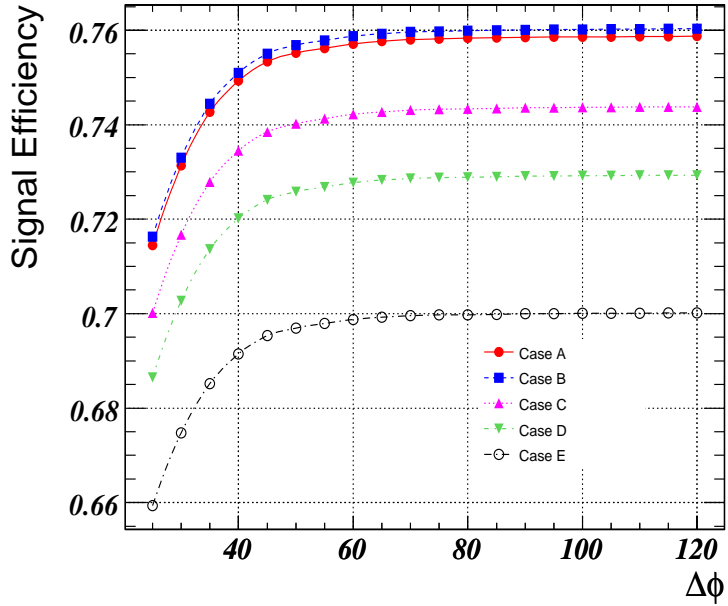}}
\caption{Signal efficiency as a function of $\Delta\phi$ variable for various selection cuts mentioned in Table~\ref{table:Cmp_Selection}.}
\label{fig:Deltaphi}
\end{center}
\end{figure}

\begin{figure}[!Hhtb]
  \begin{center}
  \resizebox{0.65\linewidth}{0.5\linewidth}{\includegraphics{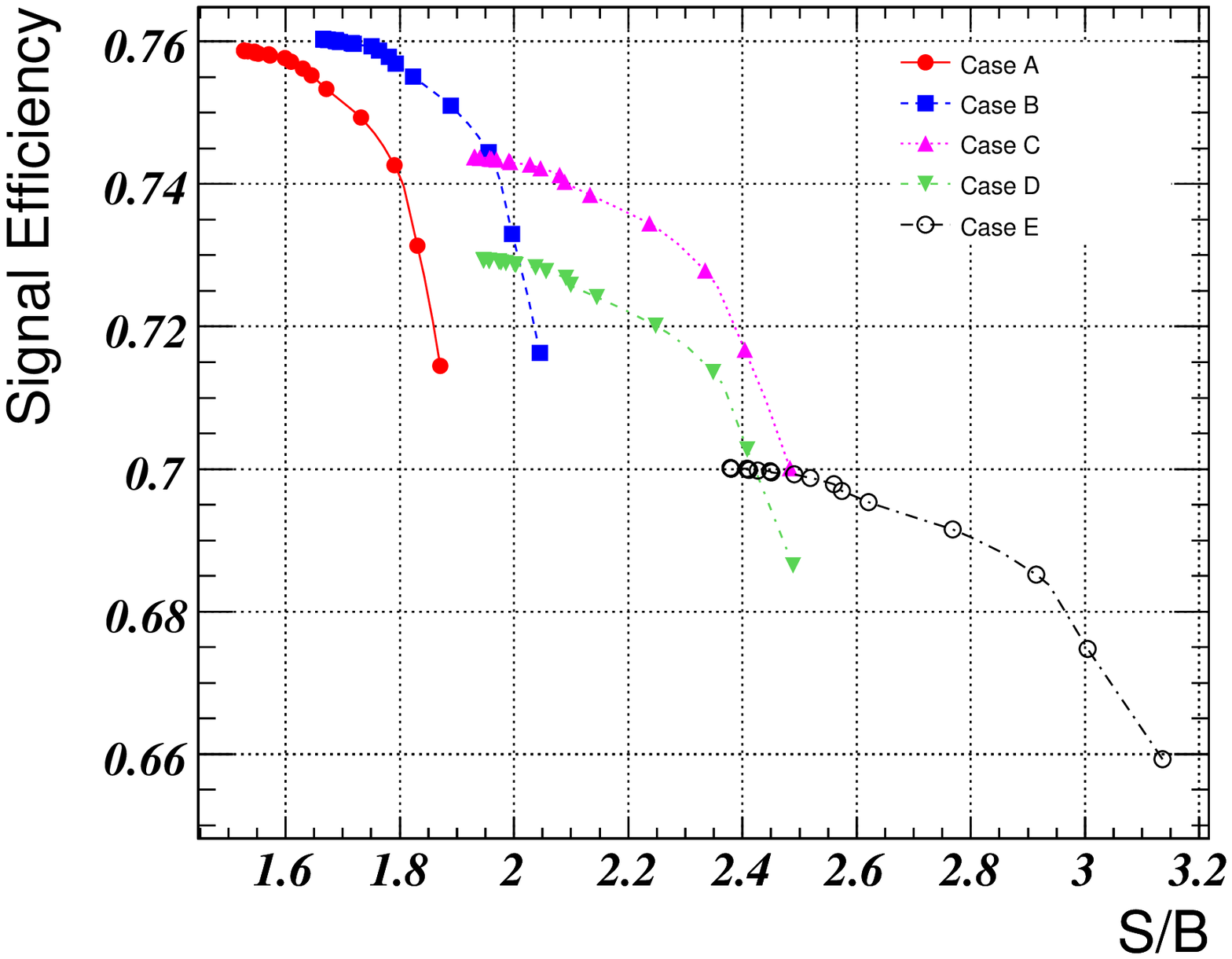}} 
\caption{Signal efficiency as a function of S/B for various selection cuts mentioned in Table~\ref{table:Cmp_Selection}. Each curve represents variation in $\Delta\phi$ for a chosen set of cut value from Table~\ref{table:Cmp_Selection}. }
\label{fig:Deltaphi1}
\end{center}
\end{figure}

In $\gamma$ + jets events, the jets have been reconstructed using Iterative Cone type algorithm with cone size R =0.5\cite{DAQTDR}. Since the HCAL extends up to $|\eta| < 5.0$ the selection cuts used on the highest $P_{T}$ jet are : $P_{T}^{jet} > 40$ GeV and $|\eta^{jet}|<5.0$. Usually $\gamma$ + 1 jet events are produced in a ``back-to-back" fashion, but this topology is disturbed by initial-state and final-state radiation. The events with the vector $P^{jet}_{T}$ being ``back-to-back" to the vector $P^{\gamma}_{T}$ within $\Delta\phi$ can be defined by the equation;                           
   \begin{center}$\Phi(\gamma,jet)$ = $180^{0}$ $\pm$ $\Delta\phi$ \end{center}
   where $\Phi(\gamma,jet)$ is the angle between $P^{\gamma}_{T}$ and $P^{jet}_{T}$ vectors. To further optimize S/B ratio while retaining a very large signal efficiency, the effect of $\Delta\phi$ cut has been studied after applying all photon isolation requirements. Fig.~\ref{fig:Deltaphi} shows the signal efficiency as a function of $\Delta\phi$ cut. Fig.~\ref{fig:Deltaphi1} shows the performance of $\Delta\phi$ as a function of signal efficiency vs. S/B ratio for all selection cuts noted in Table~\ref{table:Cmp_Selection}, when $\Delta\phi$ is varied in steps of $5^{0}$. It is found that the limit $ \Delta\phi< 40^{0}$ can be used as an additional cut to improve the S/B ratio. Requiring $ \Delta\phi< 40^{0}$ improves the S/B ratio by $\sim15\%$ with a bare reduction of $1-2\%$ in signal efficiency.   

         
\section*{Statistical and Systematic uncertainties}

The statistical uncertainty on the signal and the background rates are found to be $\sim1\%$ and $\sim5\%$ respectively. We have studied some of the potential sources of theoretical and experimental uncertainty. The uncertainty on the signal (background) rate arising from the choice of the pdf used in the event generation is found to be $\sim3\%$ ($\sim5\%$). The variation of the hard scale from default choice of $Q^{2}$ in Pythia to $Q^{2} = \hat{s}$ (MSTP 32=4) results in $\sim7\%$ uncertainty on the signal rate and $\sim27\%$ on the background rate. The estimated uncertainty due to photon trigger is $\sim1\%$\cite{Higgs,baffioni}. For $P^{jet}_{T}>60$ GeV, the uncertainty due to jet energy scale is $\sim\pm7\%$. Experimental measurement uncertainties such as  photon energy scale and jet energy resolution effect the rates within $\sim1\%$. The systematic uncertainty for the signal and the background rate arising from the preselection efficiency is less than $3\%$. These uncertainties when added in quadrature leads to $\sim11\%$ uncertainty on the signal rate and $\sim29\%$ on the background rate.

In the initial phase of the LHC operation, for an integrated luminosity of $1 fb^{-1}$, the error on the measured luminosity at CMS is expected to be $\sim10\%$\cite{PhyTDR2}. Studies from other collider experiments\cite{D0,D04,HERA} have reported uncertainties due to photon purity, inefficiency in event vertex determination and conversion of photons in the detector. However, the understanding of such sources of systematic uncertainty in the context of this analysis would be better and clear after the availablity of the LHC data.

\section*{Conclusions}
In this analysis, we have explored the direct photon + jet production at the CMS in the region of photon transverse momentum ($P^{\gamma}_{T}$) from HLT $P_{T}$ threshold of 80 GeV to a few hundred GeV. We have studied the tracker, the ECAL and the HCAL based isolation conditions for the direct photon physics. For low luminosity phase of the LHC operation, it is found that  with optimized isolation conditions the background can be reduced by two orders of magnitude while the signal efficiency remains between $70\%-80\%$. Isolation requirement within a larger cone around a photon provides a clean and pure signal. It is found that for the entire $P_{T}$ range in the analysis, carefully chosen values of the isolation threshold parameters can yield appreciable increase in the S/B ratio as compared to previous studies while retaining very high signal efficiency. We have found selection cut C (Table~\ref{table:Cmp_Selection}) to be optimum. We have matched our results from the Pythia based simulation with an independent theoretical calculation and they are in excellent agreement. The topology of the $\gamma$ + jet events is such that at high $P_{T}$ most of the  direct photon and jet events are ``back to back" within certain $\Delta\phi$. The inclusion of a very wide $\Delta\phi$ cut at $40^0$ in the analysis leads to a further increase of $\sim15\%$ thus leading to an overall gain $\sim42\%$ in S/B ratio. A better understanding of fake photon rejection will naturally lead to improved efficiency for detecting low mass Higgs decaying to photon pairs.

\section*{Acknowledgments}
We are grateful to J. Owens for providing theoretical calculations. We thank M.Pieri for helping with the background samples. We also wish to express thanks to P. Demin, D. Futyan, A. Kumar, M. Mangano, C. Seez and R. Y. Zhu. The work has been supported by a grant from the Department of Science and Technology, Govt. of India. We gratefully acknowledge the facilities provided by the Center for Detector and Related Software Technology, University of Delhi. PG would like to express gratitude to the Council of Scientific and Industrial Research, Govt.of India, for financial assistance and to Prof. R.K. Shivpuri and Prof. Raghuvir Singh for support and encouragement.



\end{document}